\documentclass [12pt] {article}
\usepackage{latexsym}
\usepackage{latexsym}
\newtheorem{theorem}{Theorem}
\newtheorem{proposition}{Proposition}
\newtheorem{corollary}{Corollary}
\newtheorem{lemma}{Lemma}

\title{Problems of robustness for universal coding schemes}
\author{V.V.V'yugin}

\date{}

\begin{document}
\maketitle

\begin{abstract}
The Lempel--Ziv universal coding scheme is asymptotically optimal for the class
of all stationary ergodic sources. A problem of robustness of this property
under small violations of ergodicity is studied. A notion of deficiency
of algorithmic randomness is used as a measure of disagreement between 
data sequence and probability measure. We prove that universal compressing 
schemes from a large class are non-robust in the following sense:
if the randomness deficiency grows arbitrarily slowly on initial fragments 
of an infinite sequence then the property of asymptotic optimality of
any universal compressing algorithm can be violated. Lempel--Ziv compressing
algorithms are robust on infinite sequences generated by ergodic 
Markov chains when the randomness deficiency of its initial fragments
of length $n$ grows as $o(n)$.
\end{abstract}

\section{Introduction}

Well known data compression schemes universal for classes of stationary
ergodic sources, like Lempel--Ziv algorithms, are asymptotically optimal 
\cite{LeZ77, LeZ78}. In particular, for almost every infinite binary sequence
$\omega_1\omega_2\dots$ generated by an ergodic source with unknown 
statistics the average length of codeword related to one bit of input sequence
tends to entropy of the source when the block length tends to infinity.
It looks significant a property of coding algorithms to be robust
under small variations of its parameters. We consider in this paper
a problem of robustness of the asymptotic optimality property under small
violations of ergodicity of a source. A notion of deficiency
of algorithmic randomness $d_P(\omega_1\dots\omega_n)$ is used 
as a measure of disagreement between data sequence 
$\omega\dots\omega_n\dots$ and probability distribution $P$. This
notion is considered in Kolmogorov theory of algorithmic complexity
and randomness \cite{Kol69, USS90, LiV97}. In the framework of this theory  
we can formulate laws of probability theory, 
i.e. statements which hold almost surely, in a ``pointwise'' form 
as statements which hold for individual objects. The set of 
Martin-L\"of~\cite{Mar66} random sequences is used at the present 
time as a standard set of such individual objects. The measure of
this set is equal 1 and laws of probability theory, like the law of large 
numbers, the law of iterated logarithm and others, hold for each sequence 
from this set. A sequence $\omega_1\omega_2\dots$ is algorithmic random
with respect to a computable measure $P$
if and only if the randomness deficiency $d_{P}(\omega_1\dots\omega_n)$ 
of its initial fragments of length $n$ is bounded
then $n$ increases (exact definition of the randomness deficiency 
will be given in Section~\ref{random-1}).

``Robustness'' under small violations of algorithmic randomness
of some probability laws was studied in \cite{Vov87, Sch71}. These
statements hold not only for random sequences but they hold 
also for sequences from more broader sets: the law of large numbers 
for symmetric Bernoulli scheme holds for any sequence 
$\omega_1\omega_2\dots$ such that 
$d_{P}(\omega_1\dots\omega_n)=o(n)$; the law of iterated logarithm
holds if $d_{P}(\omega_1\dots\omega_n)=o(\log\log n)$.
Small variations of these conditions imply violations of these laws.
Robustness property can be failed for laws of more general type.
It is proved in \cite{Vyu97} that Birkhoff's ergodic theorem is non-robust
in this sense -- any small growing of the deficiency of randomness
on initial fragments of an infinite sequence $\omega_1\omega_2\dots$
can imply the violation of the statement of this theorem.

We prove that for any unbounded, nonnegative, and nondecreasing 
function $\sigma(n)$ a stationary ergodic (and computable with respect to
$\sigma$) measure $P$ exists such that for any universal code for some
infinite binary sequence $\omega_1\dots\omega_{n}\dots$ inequality
$d_P(\omega_1\dots\omega_{n})\le\sigma(n)$ holds for all sufficiently 
large $n$ and the property of asymptotic optimality of this code
is violated for this sequence.

\section{Algorithmic complexity and randomness}\label{random-1}

Main notions and results on computability can be found in \cite{Rog72}.
In this paper we consider algorithms working with constructive
objects (that is integer and rational numbers, or words in finite
alphabet). Let $B$ be some finite alphabet and $B^*$ be the set of all
words (finite sequences of letters) in it. Empty word $\Lambda$ 
is also an element of $B^*$. Let $l(x)$ be the length (number of letters) 
of a word $x\in B^*$. We write $x\subseteq y$ if a word $x$ is a prefix 
of a word $y$. Two words $x$ and $x'$ are comparable if $x\subseteq x'$ 
or $x'\subseteq x$. Let $bx$ be a concatenation of $b$ and $x$ 
(i.e. all letters of $x$ follow after all letters of $b$ in $bx$).

Kolmogorov (algorithmic) complexity of a word $x\in B^*$
(with respect to a word $y\in B^*$) is equal to the length of the shortest
binary codeword $p$ (i.e. $p\in\{0,1\}^*$) by which given $y$ 
the word $x$ can be reconstructed
\footnote
{
We suppose that $\min\emptyset=+\infty$.
}
$$
K_\psi(x|y)=\min\{l(p):\psi(p,y)=x\}.
$$
By this definition the complexity depends on partial computable function
$\psi$ -- method of decoding. A.N.Kolmogorov proved that {\it an 
optimal decoding algorithm} $\psi$ exists such that for any
positive constant $c$ (do not depending from $x$, $y$ and $\psi'$) 
\begin{equation}\label{inv-th}
K_\psi(x|y)\le K_{\psi'}(x|y)+2K(\psi')+c
\end{equation}
holds for any computable decoding function $\psi'$ and for all words
$x$ and $y$. Here $K(\psi')$ is the length of the shortest program
computing values of $\psi'$.
\footnote
{
We suppose that some universal programming language is fixed,
and all decoding programs are written in this language
(the constant $c$ depends on this language).
}
We fix some optimal decoding function $\psi$. The value
$K(x|y)=K_\psi(x|y)$ is called (conditional) Kolmogorov complexity
of $x$ given $y$. Unconditional complexity of $x$ is defined 
$K(x)=K(x|\Lambda)$.

It follows from \cite{ZvL70} that a corresponding to $\psi$ coding algorithm
(in sense of Section~\ref{code-1}) computing by $x$ a 
codeword $p$ of minimal length such that $\psi(p)=x$ does not exist.

We will use some properties of Kolmogorov complexity \cite{LiV97, ZvL70}.
Incompressibility property asserts that for any positive integer numbers
$n$ and $m$ a portion of all sequences $x$ of length $n$ such that
\begin{equation}\label{incom-1}
K(x)<n-m,
\end{equation}
is less than $2^{-m}$. Indeed, the number of all $x$ satisfying 
this inequality does not exceed the number of all binary programs
generating them. Since the length of any such program is less than $n-m$
the number of these programs is less than $2^{n-m}$.

Let $x$ and $b$ be finite words. It is easy to construct
a function which given any program computing $bx$ and the length of $b$
computes the word $x$. Therefore,
\footnote
{
We will consider in the following logarithms on the base 2.
}
\begin{equation} \label{K-1}
K(x)\le K(bx)+2\log l(b)+c 
\end{equation}
for any $x$, where $c$ is a positive constant not depending from
$b$ and $x$.

We consider a probability space $(\Omega,F,P)$, where
$\Omega=\{0,1\}^{\infty}$, Borel field $F$ is generated by balls
$\Gamma_x=\{\omega\in\Omega : x\subseteq\omega\}$, where $x\in\{0,1\}^*$.
To define a probability measure $P$ on the space $\Omega$ 
it is sufficient to define the concordant values $P(\Gamma_x)=P(x)$ 
such that $P(\Lambda)=1$ and $P(x)=P(x0)+P(x1)$ for all
$x$, where $x\nu$ denotes a word obtained from $x$ by adding $\nu$
on right. After that, the function $P$ can be extended by Kolmogorov
extension theorem \cite{Shi80}. A uniform Bernoulli probability distribution
on binary sequences is defined $B_{1/2}(x)=2^{-l(x)}$.
A measure $P$ is called computable if there exists an algorithm
which given a finite sequence $x$ and a degree of accuracy, a
rational $\epsilon>0$, outputs a rational approximation to $P(x)$
with the accuracy $\epsilon$.

A notion of algorithmic random sequence is defined using an algorithmic
analogue of a set of measure $0$. Let $P$ be a computable probability
measure on a set of all infinite binary sequences $\Omega$. 

A set $M\subseteq\Omega$ has $P$-measure 0 if for each
rational $\epsilon>0$ there is a sequence $x(1),x(2),\dots$ of elements
of $\Xi$ such that the set $U_{\epsilon}=\cup_i\Gamma_{x(i)}$ satisfies
$M\subseteq U_{\epsilon}$ and $P(U_{\epsilon})<\epsilon$.
A $P$-null set is called effectively $P$-null if there exists a computable 
function $x(\epsilon,i)$ such that 
$M\subseteq U_{\epsilon}=\cup\Gamma_{x(\epsilon,i)}$ and
$P(U_{\epsilon})<\epsilon$ for each rational $\epsilon>0$.
It can be proved that for any computable measure $P$ there
exists the largest with respect to the measure-theoretic inclusion 
effectively $P$-null set \cite{USS90, LiV97, Mar66}.  
The complement of this largest effectively $P$-null set is called the
constructive support of the measure $P$. 
An infinite sequence $\omega\in\Omega$ is called algorithmic random 
with respect to the measure $P$ (random in the sense of Martin-L\"of) 
if it belongs to the constructive support of the measure $P$. 

Using some modification of decoding algorithms we can define a notion
of algorithmic random sequence in terms of complexity 
\cite{USS90, LiV97, Lev73}.
Let us consider monotonic computable transformations of sequences.
Let $A$ and $B$ be finite alphabets, and let a set
$\hat\psi\subseteq A^*\times B^*$ is (recursively) enumerable 
(by means of some algorithm) and such that
for any $(x,y), (x',y')\in\hat\psi$ if $x$ and $x'$ are comparable then
$y$ and $y'$ are also comparable. Let also $A=\{0,1\}$.
The set $\hat\psi$ defines some monotonic with respect to $\subseteq$
decoding function
\footnote
{
Here the by supremum we mean an union of all comparable $x$ in one sequence.
}
\begin{equation}\label{mon-cod}
\psi(p)=\sup\{x:(p,x)\in\hat{\psi}\}.
\end{equation}
The class of such monotonic functions $\psi$ determines the
corresponding algorithmic complexity
$$
Km_\psi(x)=\min\{l(p):x\subseteq\psi(p)\}.
$$
The corresponding optimal complexity $Km(x)$ is differ from 
complexity $K(x)$ by a term of order of logarithm from $l(x)$. We have
\begin{equation} \label{K-Km-1}
K(x)-2\log l(x)-c\le Km(x)\le K(x)+2\log K(x)+c
\end{equation}
for all $x$, where $c$ is a positive constant \cite{USS90, LiV97}. 

For any sequence $\omega$ denote by
$\omega^n=\omega_1\dots\omega_n$ its initial fragment of length $n$.
The following fundamental assertion (which at first was proved 
in \cite{Lev73}) holds.
\begin{proposition} \label{Km-random}
Let $P$ be some computable measure. Then

1) for any infinite sequence $\omega$ a constant $c$ exists such that
for all $n$ inequality $Km(\omega^n)\le -\log P(\omega^n)+c$ holds,
besides, for any $m$  
$$
P(\cup\{\Gamma_x: -\log P(x)-Km(x)\ge m\})\le 2^{-m};
$$

2) a sequence $\omega$ is random with respect to a measure $P$ in sense 
of Martin-L\"of if and only if for some constant $c$ it holds
$Km(\omega^n)\ge -\log P(\omega^n)-c$ for all $n$.
\end{proposition}

These proposition shows that asymptotic behaviour of the function
$$
d_P(\omega^n)=-\log P(\omega^n)-Km(\omega^n)
$$
can be used as a quantitative measure of nonrandomness of the 
sequence $\omega$. By Proposition~\ref{Km-random}
a sequence $\omega$ is algorithmic random with respect to a
computable measure $P$ if and only if $\sup\limits_n d_P(\omega^n)<\infty$.
The value $d_P(\omega^n)$ is called the {\it deficiency of algorithmic
randomness} of a word (finite sequence) $\omega^n$ with respect to a
measure $P$ \cite{USS90, LiV97, KoU87}.

Basic notions of ergodic theory can be found in \cite{Bil69} (see also 
Appendix~2 to this paper). A property of ``asymptotic optimality
of compression'' by means of the shortest codeword defining the Kolmogorov
complexity holds.
\begin{corollary} \label{S-Mc-B-1}
Let $P$ be an arbitrary computable stationary ergodic measure, and let
$H$ be its entropy. Then for $P$-almost all infinite sequences
$\omega$ the following limits exist and the corresponding equalities hold
\begin{equation} \label{SMB-1}
\lim\limits_{n\to\infty}\frac{K(\omega^n)}{n}=
\lim\limits_{n\to\infty}\frac{Km(\omega^n)}{n}=
\lim\limits_{n\to\infty}\frac{-\log P(\omega^n)}{n}=H.
\end{equation}
\end{corollary}
This corollary follows from Proposition~\ref{Km-random}, 
relation (\ref{K-Km-1}) and Shannon -- McMillan -- Breiman 
theorem~\cite{Bil69}.
At first this corollary was proved for $K(x)$ in~\cite{ZvL70}.
In \cite{Vyu98} a variant of~(\ref{SMB-1}) for algorithmic random sequence was
obtained: for any infinite sequence $\omega$ random with respect to
a computable ergodic measure $P$ with entropy $H$ relations~(\ref{SMB-1}) 
hold where the limit is replaced on upper limit.

\section{Non-robustness property of the universal data 
compression scheme}\label{non-robust-1}

It looks important a property of compressing algorithms to be robust
under small variations of its parameters.
The following Theorem~\ref{theorem-1} can be interpreted as an assertion
of that ``optimal compression scheme'' corresponding to Kolmogorov complexity
is non-robust in the class of all stationary ergodic sources.
As consequences of this theorem we obtain in Section~\ref{code-1}
results on non-robustness of computable universal coding schemes 
(see Propositions~\ref{theorem-3} and \ref{theorem-4}).

\begin{theorem}   \label{theorem-1} 
For any nonnegative, nondecreasing, 
and unbounded function $\sigma(n)$ and for any real number
$0<\epsilon<1/4$ a computable with respect to $\sigma$ stationary ergodic
measure $P$ with entropy $0<H\le\epsilon$ and an infinite binary sequence
$\alpha$ exist such that
\begin{eqnarray}\label{def-lim-1}
d_{P}(\alpha^n)\le\sigma(n)
\end{eqnarray}
for almost all $n$. It holds also
\begin{eqnarray}
\limsup\limits_{n\to\infty}\frac{K(\alpha^n)}{n}\ge\frac{1}{4}, 
\label{limsup-1}
\\
\liminf\limits_{n\to\infty}\frac{K(\alpha^n)}{n}\le\epsilon. \label{liminf-1}
\end{eqnarray}
\end{theorem}
{\it Proof}. 
Let $r>0$ be a sufficiently small rational number. Let us consider
a partition
\begin{eqnarray*}
\pi_0=[0,\frac{1}{2})\cup (\frac{1}{2}+r,1],  
\pi_1=[\frac{1}{2},\frac{1}{2}+r]
\end{eqnarray*}
of semiopen interval $[0,1)$ (the number $r$ will be specified later). 
Using cutting and stacking method
(basic definitions for this method will be given in Appendix~2) we will
define an ergodic transformation $T$ of interval $[0,1)$ which will
generate a stationary ergodic measure $P$ on the set $\Omega$.
To define the measure $P$ consider
\begin{equation} \label{mes-1}
P(a_1a_2\dots a_n)=\lambda\{\omega : 
\omega\in [0,1),\mbox T^i(\omega)\in\pi_{a_i}\mbox, i=1,2,\dots,n\},
\end{equation}
where $a_1a_2\dots a_n$ is an arbitrary binary sequence,
$\lambda$ is the uniform measure on the interval $[0,1)$.
The measure $P$ is extended on arbitrary Borel subsets of $\Omega$
by a natural fashion \cite{Shi80}.

The ergodic transformation $T$ will be defined by a sequence of gadgets
$\Delta_{s}$, $\Pi_{s}$, where $s=0,1,\dots$. 
Let a gadget $\Phi_s$ be the union of these two gadgets.
We define at step $s$ an approximation $T_s=T(\Phi_s)$ of the transformation
$T$ and corresponding approximation $P^s$ of the measure $P$ 
analogously to (\ref{mes-1}). The transformation $T_s$ determines
finite trajectories starting in the points of internal intervals
of these gadgets and finishing in the top intervals. Any such trajectory
has a name which is a word in the alphabet $\{0,1\}$.
By definition for any word $a$ (for any set of words $D$)
the number $P^s(a)$ ($P^s(D)$ accordingly) is equal to the sum
of lengths of all intervals of the gadget $\Pi_{s}$ from which
trajectories with names extending $a$ (extending words from $D$) start.

Since the function $\sigma$ is nondecreasing and unbounded
a computable with respect to it sequence of positive integer numbers exists
such that $0<h_{-2}<h_{-1}<h_0<h_1<\dots$ and
\begin{equation}\label{speed-1}
\sigma(h_{i-1})-\sigma(h_{i-2})>-\log r+i+13 
\end{equation}
for all $i=0,1,\dots$. 
The gadgets will be defined by mathematical induction on steps.
The gadget $\Delta_{0}$ is defined by cutting of the interval
$[\frac{1}{2}-r,\frac{1}{2}+r)$ on $2h_0$ equal parts and by stacking them.
Let $\Pi_0$ be a gadget defined by cutting of intervals
$[0,\frac{1}{2}-r)$ and $(\frac{1}{2}+r,1]$ in $2h_0$ equal parts
and stacking them. The purpose of this definition is to construct initial
gadgets of height $2h_0$ with supports satisfying
$\lambda(\hat\Delta_0)=2r$ and $\lambda(\hat\Pi_0)=1-2r$.

The sequence of gadgets $\{\Delta_{s}\}$, $s=0,1,\dots$, will
define an approximation of the uniform Bernouli measure concentrated on
the names ot their trajectories. The sequence of gadgets
$\{\Pi_s\}$, $s=0,1,\dots$, will define a measure with sufficiently small
entropy. The gadget $\Pi_{s-1}$ will be extended at each step of the 
construction by a half part of the gadget $\Delta_{s-1}$. After that, the 
independent cutting and stacking process will be applied to this extended
gadget. This process eventually defines infinite trajectories
of points from interval $[0,1)$. The sequence of gadgets 
$\{\Pi_s\}$, $s=0,1,\dots$, will be complete and will define the needed
measure $P$. Lemmas~\ref{well-def} and \ref{M-fold} will ensure
the transformation $T$ and measure $P$ to be ergodic.

The purpose of the construction is to suggest conditions under which 
there exists a point in interval $[0,1)$ having an infinite trajectory
with a name $\alpha$ satisfying~(\ref{def-lim-1}), (\ref{limsup-1}) 
and (\ref{liminf-1}). To implement~(\ref{limsup-1}) we periodically
extend initial fragments of $\alpha$ by names of trajectories
of gadgets $\Delta_{s-1}$ (for suitable $s$) which have the maximal
complexity. To bound the deficiency of randomness of initial fragment
of length $n$ by the value $\sigma(n)$ we suggest with the help of
condition~(\ref{speed-1}) some relation between
the height of the gadget $\Delta_{s}$ and the measure of the support of
this gadget. We will use Proposition~\ref{deff-1}
to define an extension with sufficiently small deficiency of randomness.
To implement condition~(\ref{liminf-1}) it is sufficient to extend
names in long runs of the construction only in account of trajectories 
of gadgets $\{\Pi_s\}$, $s=0,1,\dots$. For any $s$ only a portion
$\le r$ of the support of such gadget belongs to element $\pi_1$
of the partition. Then by ergodic theorem the most part of
(sufficiently long) trajectories of this gadget will visit $\pi_1$
according to this frequency, and the names of these trajectories
will have the frequency of ones bounded by a small number $2r$,
that ensures the bound (\ref{liminf-1}).

{\it Construction}. 
Let at step $s-1$ ($s>0$) gadgets $\Delta_{s-1}$ and $\Pi_{s-1}$
were defined. Cut of the gadget $\Delta_{s-1}$ into two copies 
$\Delta'$ и $\Delta''$ of equal width (i.e. we cut of each column into two 
subcolumns of equal width) and join $\Pi_{s-1}\cup\Delta''$ 
in one gadget. Find a number $R_s$ and do $R_s$-fold independent cutting 
and stacking of the gadget $\Pi_{s-1}\cup\Delta''$ and also of the 
gadget $\Delta'$ to obtain new gadgets $\Pi_s$ and $\Delta_{s}$ of height 
$2h_s$ such that the gadget $\Pi_{s-1}\cup\Delta^{''}$ is 
$(1-1/s)$--well--distributed in the gadget $\Pi_s$. The needed
number $R_s$ exists by Lemma~\ref{M-fold} (Appendix~2). 

{\it Properties of the construction}. Define $T=T\{\Pi_s\}$.
Since the sequence of the gadgets $\{\Pi_s\}$ is complete
(i.e. $\lambda({\hat\Pi}_s)\to 1$ and $w(\Pi_s)\to 0$
as $s\to\infty$) the transformation $T$ is defined for $\lambda$-almost 
all $\omega$. The measure $P$ is defined by (\ref{mes-1}).
The measure $P$ is stationary, since the transformation $T$ preserves
the uniform measure $\lambda$. Measure $P$ is ergodic by Lemma~\ref{well-def} 
(Appendix~2), where $\Upsilon_s=\Pi_{s}$, since the sequence of gadgets
$\Pi_s$ is complete. Besides, the gadget $\Pi_{s-1}\cup\Delta''$, 
and the gadget $\Pi_{s-1}$ are $(1-1/s)$~--~well--distributed 
in $\Pi_s$ for any $s$. By construction
\begin{eqnarray}\label{val-1}
\lambda(\hat\Delta_i)=2^{-i+1}r \quad \mbox{\rm and} \quad  
\lambda(\hat\Pi_i)=1-2^{-i+1}r
\end{eqnarray}
for all $i=0,1,\dots$.

This construction is algorithmic effective, so the measure $P$ is computable
with respect to $\sigma$.

Let us prove that entropy $H$ of the measure $P$ do not exceed $\epsilon$.
Since $\lambda(\pi_1)=r$ and the transformation $T$ preserves the measure
$\lambda$, by ergodic theorem in almost all points of interval
$[0,1)$ a trajectory starts such that the limit of the frequency
of visiting the element $\pi_1$ by this trajectory is equal $r$, 
when the length of initial fragment of such trajectory tends to infinity.
\footnote
{
For any $\omega\in [0,1)$ the frequency of visiting of $\pi_1$ 
by trajectory starting in $\omega$ is equal to
$(1/l)\sum_{i=1}^l\chi_1(T^i\omega)$, where $l$ is the length of this
trajectory and $\chi_1(r)=1$ if $r\in\pi_1$, and $\chi_1(r)=0$, otherwise.
}
Thus for any $\delta>0$ for all sufficiently large $n$ the measure $P$
of all sequences $x$ of length $n$ with portion of ones $\le 2r$
is $\ge 1-\delta$. Let us consider any such sequence $x$ as an element
a finite set consisting of all sequences of length $n$ and containing
no more than $2rn\le\frac{n}{2}$ ones. Then we obtain a standard upper
bound
\begin{equation} \label{cnk-1}
\frac{K(x)}{n}\le\frac{1}{n}\log\left(2rn{n \choose 2rn}\right)+
\frac{2\log n}{n}\le -3r\log r
\end{equation}
for all sufficiently large $n$. By this inequality and by (\ref{SMB-1})
we obtain upper bound $H\le -3r\log r\le\epsilon$ for entropy $H$ of the
measure $P$, where $r$ is sufficiently small.

Let us prove that an infinite sequence $\alpha$ exists such that
the conclusion of Theorem~\ref{theorem-1} holds. We will define $\alpha$
by induction on steps~$s$ as the union of an increasing sequence of 
initial fragments  
\begin{equation} \label{alpha-1}
\alpha(0)\subset\dots\subset\alpha(k)\subset\dots
\end{equation}
For all sufficiently large $k$ the Kolmogorov complexity
of initial fragment $\alpha(k)$ will be small if $k$ is odd,
and complexity of $\alpha(k)$ will be large, otherwise.

Define $\alpha(0)$ be equal to $\Pi_0$--name of some trajectory of 
length $\ge h_0$ such that $d_P(\alpha(0))\le 2$. 
This is possible to do by Proposition~\ref{deff-1} (Appendix~1). Define 
$s(-1)=s(0)=0$.

{\it Induction hypotheses.} Suppose that $k>0$ and a sequence
$\alpha(0)\subset\dots\subset\alpha(k-1)$ is already defined, 
and for some step $s(k-1)$ of the construction the word $\alpha(k-1)$ 
is $\Pi_{s(k-1)}$~--~name of a trajectory of some point from the support 
of the gadget $\Pi_{s(k-1)}$. We suppose that
$l(\alpha(k-1))>h_{s(k-1)}$, and if $k$ is odd then 
$d_P(\alpha(k-1))\le\sigma(h_{s(k-2)})-4$. 
If $k$ is even then $d_P(\alpha(k-1))\le\sigma(h_{s(k-2)})$ and
$P^{s(k-1)}(\alpha(k-1))>(1/8)P(\alpha(k-1))$.

Let us consider any odd $k$. Define $a=\alpha(k-1)$.

Let us consider a set of all intervals (from columns) of the 
gadget $\Pi_{s-1}$ with the following property: for any trajectory 
starting from this interval with $\Pi_{s-1}$-names extending $a$
the frequency  of visiting the element $\pi_1$ of the partition is $\le 2r$.
For the name $\gamma$ of any such trajectory an inequality
\begin{equation} \label{small-1}
K(\gamma)/l(\gamma)\le -3r\log r\le\epsilon
\end{equation}
(analogous to (\ref{cnk-1})) holds, where $r$ is sufficiently small. 
As in the proof of the inequality $H\le\epsilon$ we obtain by ergodic
theorem that for all sufficiently large $s$ total length of all interval
from this set is $\ge (1/2)P(a)$.

Let us consider an arbitrary column from the gadget $\Pi_s$.
Divide all its intervals on two equal parts: upper part and lower part.
We will consider only intervals from the lower part. Any trajectory starting
from a point of an interval from this part has length $\ge h_s$.
Fix some $s$ as above and define $s(k)=s$.
Let $U_{s}(a)$ be all intervals from the lower part of the gadget 
$\Pi_{s}$ such that trajectories starting from them and having 
$\Pi_{s}$~--~names extending $a$ satisfy the inequality~(\ref{small-1}). 
Let $D_a$ be a set of all $\Pi_{s}$~--~names of all these trajectories.
Inequality $P^s(D_a)=P^s(a)>(1/4)P(a)$ holds for the total length 
$P^s(D_a)$ of all intervals from $U_{s}(a)$.

Define $\tilde D=\cup_{x\in D}\Gamma_x$.
It is easy to prove that a set $C_a\subseteq D_a$ exists such that
$P(\tilde C_a)>(1/8)P(\tilde D_a)$ 
and $P^s(b)>(1/8)P(b)$ for all $b\in C_a$. 
By Proposition~\ref{deff-1} (Appendix~1) an $b\in C_a$ exists such that
$d_P(b^j)\le d_P(a)+4$ when $l(a)\le j\le l(b)$. Define $\alpha(k)=b$.
By induction hypotheses inequalities
$d_P(a)\le\sigma(h_{s(k-2)})-4$ and $l(a)\ge h_{s(k-1)}>h_{s(k-2)}$ hold.
Then $d_P(b^j)\le\sigma(h_{s(k-2)})\le\sigma(l(a))\le\sigma(j)$ for all
$l(a)\le j\le l(b)$.

Notice, that $l(b)\ge h_{s(k)}$, since any trajectory defining
$b$ starts from an interval of the lower part of the gadget $\Pi_s$,
and the height of this gadget is $\ge 2h_s$.
The rest induction hypotheses are proved above.

The condition (\ref{liminf-1}) is true, since condition~(\ref{small-1})
holds for infinite number of initial fragments $\alpha(k)$ 
of the sequence $\alpha$.

Let $k$ be even. Put $b=\alpha(k-1)$.
Let $s=s(k-1)+1$. Define $s(k)=s$. 

Let us consider an arbitrary column from the gadget $\Delta_{s-1}$.
Divide all its intervals into two equal parts: upper part and lower part.
Any trajectory starting from an interval of the lower part have 
the length $\ge L/2$, where 
$L\ge 2h_{s-1}$ is the height of the gadget $\Delta_{s-1}$.
The uniform measure of all such intervals is equal to
$\frac{1}{2}\lambda(\hat\Delta_{s-1})$.
Let us consider the names $x^{L/2}$ of initial fragments of length
$L/2$ of all these trajectories. By incompressibility property
of Kolmogorov complexity (\ref{incom-1}) and by choice of
$L$ the uniform Bernoulli measure of all sequences of length
$L/2$ satisfying
$$
\frac{K(x^{L/2})}{l(x^{L/2})}<1-\frac{2}{h_{s-2}},
$$
is less than $2^{-L/h_{s-2}}\le 1/4$. Names of initial fragments 
(of length $L/2$) of the rest part of trajectories starting from
intervals of lower part of the gadget $\Delta_{s-1}$ satisfy
\begin{equation} \label{K-2}
\frac{K(x^{L/2})}{l(x^{L/2})}\ge 1-\frac{2}{h_{s-2}}.
\end{equation}
It is noted in Appendix~2 (Remark~1), for any step $s$ of the 
construction the equality
$P^{s-1}(x)=2^{-l(x)}\lambda(\hat\Delta_{s-1})$ holds for the name
$x$ of any trajectory of the gadget $\Delta_{s-1}$. We conclude
from this equality that the uniform measure of all intervals
from the lower part of the gadget $\Delta_{s-1}$, such that
trajectories with names (more correctly, with initial fragments $x^{L/2}$
of such names) satisfying (\ref{K-2}) start from these
intervals, is at least $\frac{1}{4}\lambda(\hat\Delta_{s-1})$.

By~(\ref{speed-1}) and (\ref{val-1})
\begin{eqnarray}\label{vol-2}
\gamma=\frac{\lambda(\hat\Delta'')}{\lambda(\hat\Pi_{s-1})}=
\frac{\lambda(\hat\Delta_{s-1})}{2\lambda(\hat\Pi_{s-1})}
=\frac{2^{-s+1}r}{1-2^{-s+2}}>
2^{-s+1}r\ge 2^{-(\sigma(h_{s-1})-\sigma(h_{s-2})+12}
\end{eqnarray}
Let us consider $R_{s}$--fold independent cutting and stacking
of the gadget $\Pi_{s-1}\cup\Delta''$ in more details. At first,
we cut of this gadget on $R_{s}$ copies. When we stack the next copy on
already defined part of the gadget the portion of all trajectories
of any column from the previously constructed part, which go to a
subcolumn from the gadget $\Delta''$, is equal to
\begin{equation} \label{ratio-1}
\frac{\lambda(\hat\Delta'')}{\lambda(\hat\Pi_{s-1})+\lambda(\hat\Delta'')}
=\frac{\gamma}{1+\gamma}.
\end{equation}
This is true, since by definition any column is covered by a set of 
subcolumns with the same distribution 
as the gadget $\Pi_{s-1}\cup\Delta''$ has. Total length
of all intervals of the gadget $\Pi_{s-1}$ such that trajectories 
with names extending $b$ start from these intervals is equal to
$P^{s-1}(b)$. 

Consider the lower half of all subintervals generated by cutting and
stacking of the gadget $\Pi_{s-1}$ in which trajectories with
$\Pi_{s-1}$--names extending $b$ start. The length of any such 
trajectory (in $\Pi_{s}$) is at least $h_s$. By this reason
some inductive hypothesis will be true. The measure of all
remaining subintervals decreases twice. After that, we consider
a subset of these subintervals, such that trajectories starting
from subintervals of this subset go into subcolumns of the gadget 
$\Delta''$. The measure of remaining subintervals is multiplied
by a factor $\gamma/(1+\gamma)$. Further, consider subintervals
from the remaining part generating trajectories whose names have in 
$\Delta''$ fragments satisfying~(\ref{K-2}). The measure of
the remaining part can be at least $1/4$ from the previously considered
part. We obtain this bound from previous estimate of the portion
of subintervals generating trajectories in the gadget $\Delta''$ 
of length $\ge L/2$ satisfying~(\ref{K-2}).
\footnote
{
Remember, that $L$ ($\ge 2h_{s-1}$) is the height of gadgets 
$\Pi_{s-1}$, $\Delta_{s-1}$.
}
Let $D_b$ be a set of all $\Pi_s$--names of all trajectories starting from
subintervals remaining after these selection operations. Then
\begin{equation}\label{inequ-1}
P^s(D_b)\ge\frac{\gamma}{8(1+\gamma)}P^{s-1}(b).
\end{equation}
The name of any such trajectory has initial fragment of type
$bx'x^{L/2}$, where $x'x^{L/2}$ is the name of a fragment 
of this trajectory corresponding to its path in the gadget $\Delta_{s-1}$.
The word $x^{L/2}$ has length $L/2$ and satisfies (\ref{K-2}).
The word $x'$ is the name of a fragment of the trajectory which goes
from lower interval to an interval generating trajectory with name $x^{L/2}$. 
We have $l(bx'x^{L/2})\le 2L=4l(x^{L/2})$. By (\ref{K-1}) and (\ref{K-2}) 
we obtain for these initial fragments of sufficiently large length
\begin{equation} \label{K-3}
\frac{K(bx'x^{L/2})}{l(bx'x^{L/2})}\ge\frac{K(x^{L/2})-
2\log l(bx')}{4l(x^{L/2})}\ge\frac{1}{4}-\frac{1}{h_{s-2}}.
\end{equation}
We have $P^{s-1}(b)>(1/8)P(b)$ by induction hypothesis.
After that, taking into account that $\gamma\le 1$, we deduce from
(\ref{inequ-1})
\begin{eqnarray*}
P(\tilde D_b)\ge P^{s-1}(D_b)\ge\frac{\gamma}{128}P(b).
\end{eqnarray*}
By Proposition~\ref{deff-1} an $c\in D_b$ exists such that
\begin{eqnarray*}
d_{P}(c^j)\le d_{P}(b)+1-\log\frac{\gamma}{128}\le    \\
d_{P}(b)+(\sigma (h_{s-1})-\sigma (h_{s-2})-12)+8\le
\sigma (h_{s-1})-4=\sigma (h_{s(k-1)})-4
\end{eqnarray*}
for all $l(b)\le j\le l(c)$. Here we have
$d_P(b)\le\sigma (h_{s(k-2)})\le\sigma(h_{s-2})$ by induction hypothesis. 
We also used inequality (\ref{vol-2}). Besides, by induction hypothesis
we have $l(b)\ge h_{s-1}$. Therefore,
$$
d_{P}(c^j)<\sigma(h_{s-1})\le\sigma(l(b))\le\sigma(j)
$$ 
for $l(b)\le j\le l(c)$. Define $\alpha(k)=c$. 
It is easy to see that all induction hypotheses are true for $\alpha(k)$.

An infinite sequence $\alpha$ is defined by a sequence of initial
fragments (\ref{alpha-1}). We proved that 
$d_{P}(\alpha^j)\le\sigma(j)$ for all $j\ge l(\alpha(1))$.

By the construction there are infinitely many initial fragments
of the sequence $\alpha$ satisfying~(\ref{K-3}). The sequence $h_s$, 
where $s=0,1,\dots$, is monotone increased. So, the
condition~(\ref{limsup-1}) hold. 
$\bigtriangleup$

\section{Non-robustness property of universal codes} \label{code-1}

Let $A$ and $B$ be finite alphabets. By {\it a code} we mean
a computable family of functions
\footnote
{
A function $\phi_n(\alpha)$ is computable by both arguments 
$n$ and $\alpha$.
}
$\phi_n:A^n\to B^*$, where $n=1,2,\dots$. Suppose that $B=\{0,1\}$.
We will consider decodable codes. A computable family of decoding 
functions $\psi_n:\phi_n(A^n)\to A^n$ such that
$\alpha =\psi_n(\phi_n(\alpha))$ for all $n$ and for all $\alpha\in A^n$
is associated with this code. A separating property of the code is required.
An algorithm must exist decoding any sequence of concatenated codewords.
Prefix codes satisfy to this requirement. Any two codewords
$\phi_n(\alpha)$ and $\phi_n(\alpha')$ are incomparable under prefix
method of coding. For any code $\{\phi_n\}$ a compressing ratio
$\rho_{\phi_n}(\alpha^{n})=l(\phi_n(\alpha^{n}))/(n\log|A|)$
of input word $\alpha^{n}\in A^n$ is defined. We suppose for simplicity
that $A=\{0,1\}$.

In \cite{Fit66, Dav73} codes universal in the mean for some classes of sources
were considered, in \cite{LeZ77, LeZ78} a code universal almost everywhere 
for the class of all stationary ergodic sources was defined.
We consider codes universal almost everywhere.

A code $\{\phi_n\}$ is called {\it universal} with respect to a class 
of stationary ergodic sources if for any computable stationary ergodic 
measure $P$ from this class
\begin{equation}\label{asym-zv}
\lim_{n\to\infty}\rho_{\phi_n}(\omega^n)=H
\end{equation}
holds $P$--almost every infinite sequence
$\omega=\omega_1\omega_2\dots$, where $H$ is the entropy of the measure $P$.
There exist several types of Lempel - Ziv universal coding scheme 
\cite{LeZ77, LeZ78}. Let us recall two of them.

A coding algorithm is fed with a word $\omega_1\dots\omega_N$ of length $N$.
By the first variant of the algorithm a sequence of letters
$\omega_1,\omega_2\dots\omega_n$ is read beginning at the left and
is divided on subblocks as follows: a pointer on $k$-th subblock
is inserted after $\omega_{i(k)}$ if subblock 
$\omega_{i(k-1)+1}\omega_{i(k-1)+2}\dots\omega_{i(k)-1}$ was already seen
between previous pointers and subblock
$\omega_{i(k-1)+1}\omega_{i(k-1)+2}\dots\omega_{i(k)}$ was not seen.
To encode new subblock it is sufficient to memorize coordinate
of the beginning of the sequence
$\omega_{i(k-1)+1}\omega_{i(k-1)+2}\dots\omega_{i(k)-1}$, its length,
and new letter $\omega_{i(k)}$.

The same idea is used in the second variant of the algorithm 
but a subblock
$\omega_{i(k-1)+1}\omega_{i(k-1)+2}\dots\omega_{i(k)-1}$ 
is deemed to have appeared if it occurs at all -- not necessary
between pointers.

The following proposition on non-robustness of universal
codes is an analog of Theorem~\ref{theorem-1}.

\begin{proposition}\label{theorem-3}
For any nonnegative, nondecreasing, and unbounded function $\sigma(n)$ 
and for any real number $0<\epsilon<1/4$ a computable with respect to 
$\sigma$ stationary ergodic measure $P$ with entropy $0<H\le\epsilon$ 
exists such that for each universal (for class of all stationary 
ergodic sources) code $\{\phi_n\}$ an infinite binary sequence $\alpha$
exists such that $d_{P}(\alpha^n)\le\sigma(n)$ for almost all $n$ and
\begin{eqnarray}
\limsup\limits_{n\to\infty}\rho_{\phi_n}(\alpha^{n})\ge\frac{1}{4};
\label{limsup-1mb}
\\
\liminf\limits_{n\to\infty}\rho_{\phi_n}(\alpha^{n})\le\epsilon.
\label{liminf-1mb}
\end{eqnarray}
\end{proposition}
{\it Proof}. 
For any $n$ a decoding algorithm $\psi_n$ of the code $\{\phi_n\}$ 
is defined by $\log n+O(1)$ bits. Then we have
\begin{equation}\label{K-phi-1}
K(\alpha^n)\le l(\phi_n(\alpha))+O(\log n).
\end{equation}
Inequality (\ref{limsup-1mb}) follows from the inequality
(\ref{limsup-1}) of Theorem~\ref{theorem-1}. The proof of the 
inequality~(\ref{liminf-1mb}) is analogous to the proof of the 
inequality~(\ref{liminf-1}) of Theorem~\ref{theorem-1}. 
We must only replace condition~(\ref{small-1}) from the proof
of Theorem~\ref{theorem-1} on $l(\phi_n(\omega^n))/n\le\epsilon$ 
and take into account property~(\ref{asym-zv}) of asymptotic optimality
of the code $\{\phi_n\}$.
$\bigtriangleup$

Let $\{\phi_N\}$ be a code. Under block realization of the code any
sequence of letters $\omega^n=\omega_1\dots\omega_n$ is divided
in consecutive blocks $\omega=\tilde\omega_1\dots\tilde\omega_k$, where
$n=(k-1)N+q$, $0\le q<N$ and $\tilde\omega_i=\omega_{(i-1)N}\dots\omega_{iN}$, 
$i=1,2,\dots k-1$, is a block of length $N$, and
$\tilde\omega_k=\omega_{(k-1)N}\dots\omega_{(k-1)N+q}$ is the last incomplete 
block. Any block $\tilde\omega_i$ is encoded by a binary word
$\phi_N(\tilde\omega_i)$. In asymptotic estimates (when $n\to\infty$) 
method of coding of this last block $\tilde\omega_k$ is unessential
(we fix some of these methods). We write 
$\phi_N(\omega^n)=\phi_N(\tilde\omega_1)\dots\phi_N(\tilde\omega_k)$ and
$\rho_{\phi_N}(\omega^n)=l(\phi_N(\omega^n))/n$.

It is proved in \cite{LeZ78} (Theorem 4) that for any stationary ergodic 
measure $P$ with entropy $H$ a property of asymptotic optimality holds
for block realization of Lempel--Ziv code $\{\phi_N\}$ with blocks 
of length $N$. Relation 
\begin{equation}\label{asym-zv-1}
\lim\limits_{N\to\infty}\limsup_{n\to\infty}\rho_{\phi_N}(\omega^n)=H
\end{equation}
holds for $P$--almost all $\omega$.
We can prove that equality~(\ref{asym-zv-1}) holds also for any sequence
$\omega$ random in sense of Martin-L\"of with respect to a measure $P$
(i.e. when $d_P(\omega^n)=O(1)$ as $n\to\infty$).

The following analogue of Theorem~\ref{theorem-1} holds for
block realization of codes with block length $N$ and for codes using sliding 
window of length $N$ (when a new letter of codeword depends only from 
$N$ preceding letters of input word).

\begin{proposition}\label{theorem-4}
For any nonnegative, nondecreasing, and unbounded function $\sigma(n)$ 
and for any real number $0<\epsilon<1/4$ a computable with respect to 
$\sigma$ stationary ergodic measure $P$ with entropy $0<H\le\epsilon$ 
exists such that for each universal (for class of all stationary 
ergodic sources) code $\{\phi_N\}$ or for each universal code with sliding 
window of length $N$ an infinite binary sequence $\alpha$
exists such that $d_{P}(\alpha^n)\le\sigma(n)$ for almost all $n$ and
for any $N$ 
\begin{eqnarray}
\limsup\limits_{n\to\infty}\rho_{\phi_N}(\alpha^{n})\ge\frac{1}{4},
\label{limsup-1mbb}
\end{eqnarray}
and for all sufficiently large $N$ 
\begin{eqnarray}
\liminf\limits_{n\to\infty}\rho(\phi_N(\alpha^n))\le\epsilon.
\label{liminf-1mbb}
\end{eqnarray}
\end{proposition}
The proof of this proposition is a small comlication of the proof
of Proposition~\ref{theorem-3}.

Notice, that the property~(\ref{limsup-1mbb}) is also hold
for adaptive coding scheme, i.e. when coding algorithm depends
on preceding blocks.

Using Theorem~\ref{theorem-1} it can be proved that non-robustness
property holds for other well-known universal codes.
For example, in \cite{Ryb88} a universal forecasting measure
$\rho(\omega_1\dots\omega_n)$ and a code $\psi_n$ such that
$l(\psi_n(\omega_1\dots\omega_n))\le -\log\rho(\omega_1\dots\omega_n)+1$
were defined. This measure is defined as a mixture
$\rho(y)=\sum\limits_{k=0}^{\infty}\lambda_k\rho_k(y)$ of measures
$\rho_k$ universal for Markov sources of order $k$ constructed in
the theory of universal coding \cite{KrT81}. 
Here $\lambda_k$ is some optimal probability distribution on
positive integer numbers (it can be defined
$\lambda_k=ck^{-1}\log^{-2}k$, where $c$ is a constant) 
and $\phi(k)$ is the corresponding codeword for a positive integer 
number $k$: $l(\phi(k))=\log k+O(\log\log k)$.
In \cite{Ryb84} an universal code was constructed
$\psi(u)=\phi(l(u))\psi_{l(u)}(u)$, where $u\in B^*$. 
The universality conditions for the measure $\rho$ and for the code $\psi$ 
is the following:
\footnote
{
We give some simplification of the results of \cite{Ryb88, Ryb84}.
}
for any stationary measure $\mu$ with entropy $H(\mu)$ 
for $\mu$--almost all $\omega\in\Omega$ the mean error of the forecast
by measure $\mu$ tends to zero
\begin{eqnarray}\label{univ-stab}
\lim\limits_{T\to\infty}\frac{1}{T}
\sum\limits_{t=1}^{T}\log\frac{\mu(\omega_{t+1}|\omega_1\dots\omega_t)}
{\rho(\omega_{t+1}|\omega_1\dots\omega_t)}=\lim\limits_{t\to\infty}\frac{1}{t}
\log\frac{\mu(\omega_1\dots\omega_{t})}{\rho(\omega_1\dots\omega_{t})}=0,
\end{eqnarray}
and $\lim_{n\to\infty}l(\psi(\omega^n))/n=
\lim_{n\to\infty}-\log\rho(\omega^n)/n=H(\mu)$.
It is easy to derive from the definition of the deficiency of randomness
that the condition~(\ref{univ-stab}) is ``robust under violation
of randomness'', more correctly, it holds for any computable stationary
measure $\mu$ and for any infinite sequence $\omega$ such that
$d_\mu(\omega^n)=o(n)$ as $n\to\infty$. But the corresponding
universal code $\psi$ is non-robust for the class of all stationary
ergodic sources. Since a decoding algorithm exists for the code $\psi$
it holds $K(\omega_1\dots\omega_n)\le l(\psi(\omega_1\dots\omega_n))+O(1)
\le -\log\rho(\omega_1\dots\omega_n)+O(\log n)$.
Then by Proposition~\ref{theorem-3} there exists an $\alpha\in\Omega$, 
such that the conclusion of this proposition holds, in particular,
the condition (\ref{limsup-1mb}) holds. The property~(\ref{liminf-1mb})
can be obtained as in the proof of Proposition~\ref{theorem-3}
by universality of the code.

The property of asymptotic optimality can be robust for more narrow
classes of stationary ergodic sources such that as i.i.d sequences
of random variables or stationary Markov chains. 
\begin{proposition}
Let $P$ be an arbitrary computable probability measure representing
a stationary ergodic Markov chain of fixed order (in particular,
i.i.d sequence of random variables), $H$ is its entropy, $\{\phi_n\}$
is a variant of Lempel--Ziv compressing algorithm. Then for any
infinite sequence $\omega$ if $d_P(\omega^n)=o(n)$ then
equality (\ref{asym-zv}) holds, and for block realization of this
compressing scheme equality (\ref{asym-zv-1}) holds.
\end{proposition}
The proof is based on constructive feature of the proof of results
from \cite{LeZ78}. The Birghoff's ergodic theorem is also used in this
proof that is in the case of Markov sources is a variant of
the law of large numbers. This law holds for individual sequence
$\omega$ when $d_P(\omega^n)=o(n)$ as $n\to\infty$.

{\small

\section{Appendix~1}

{\bf Bounded increase of the deficiency of randomness}.
In the proof of Theorem~\ref{theorem-1} a proposition on
a bounded increase of the deficiency of randomness was used.
Let $P$ be a measure, $P(x)\not = 0$ and a set $A$ consists of words $y$ 
such that $x\subseteq y$. Recall, that
$P(\tilde A)=P(\cup\{\Gamma_y:y\in A\})$ for any $A\subseteq\{0,1\}^*$.
Define $P(\tilde A|x)=P(\tilde A)/P(x)$.
\begin{proposition}  \label{deff-1}
Let $P$ be a measure, $x$ be a word, $P(x)\not = 0$ and a set $A$ consists 
of words $y$ such that $x\subseteq y$ and $P(\tilde A)>0$. Then for any
$0<\mu<1$ a subset $A'\subseteq A$ exists such that 
$P(\tilde A')>\mu P(\tilde A)$ and
$$
d_P(y^n)\le d_P(x)-\log(1-\mu)-\log P(\tilde A|x)
$$
for all $y\in A'$ and $l(x)\le n\le l(y)$.
\end{proposition}
{\it Proof}. We will use in the proof a notion of supermartingale 
\cite{Shi80}. 
A function $M$ is called $P$--supermartingale if it is defined 
on $\{0,1\}^*$ and satisfies conditions:

\qquad $M(\Lambda)\leq 1$;

\qquad $M(x)\geq M(x0)P(0|x)+M(x1)P(1|x)$ for all $x$,
\\where $P(\nu|x)=P(x\nu)/P(x)$ for $\nu=0, 1$ (we put here
$0/0=0*\infty=0$).

A supermartingale $M$ is lower semicomputable if the set $\{(r,x):r<M(x)\}$,
where $r$ is a rational number, is a range of some computable function.
We will consider only nonnegative supermartingales. 

Let us prove that the deficiency of randomness is bounded by a logarithm
of some lower semicomputable supermartingale.
\begin{lemma} Let $P$ be a computable probability measure. Then there
exists a lower semicomputable $P$--supermartingale $M$ such that
$d_P(x)\le\log M(x)$ for all $x$.
\end{lemma}
{\it Proof}. Let some optimal function $\psi$ satisfying
(\ref{mon-cod}) defines the monotone complexity $Km(x)$. Define
\begin{equation} \label{semi-1}
Q(\alpha)=B_{1/2}(\cup\{\Gamma_p:\alpha\subseteq\psi(p)\}),
\end{equation}
where $B_{1/2}(\Gamma_\alpha)=2^{-l(\alpha)}$ is the uniform Bernoulli measure
on the set of all binary sequences. It is easy to verify that 
$Q(\Lambda)\le 1$ and $Q(\alpha)\ge Q(\alpha0)+Q(\alpha1)$ 
for all words $\alpha$. Then the function $M(\alpha)=Q(\alpha)/P(\alpha)$ 
is a $P$--supermartingale.

Since for any $\alpha$ the shortest $p$ such that
$\alpha\subseteq\psi(p)$ is an element of the set from~(\ref{semi-1}), 
we have inequality $Q(\alpha)\ge 2^{-Km(\alpha)}$, and so,
$d_P(\alpha)\le\log M(\alpha)$.
$\bigtriangleup$

Let $d_P(x)\le\log M(x)$, where $M$ is lower semicomputable  
$P$~--~supermartingal. Let us define a set
$$
A_1=\left\{y\in A : \exists j \left(l(x)\le j\le l(y)\mbox{ and }\mbox 
M(y^j)>\frac{1}{(1-\mu)P(A|x)}M(x)\right)\right\}.
$$
A set of words $B$ is called prefix free if for any two distinct
words $x,y\in B$ conditions $x\not\subseteq y$ and $y\not\subseteq x$ hold.

By definition of supermartingale for any prefix free set $B$ 
such that $x\subseteq y$ for all $y\in B$ inequality
\begin{equation}  \label{super-1}
M(x)\ge\sum\limits_{y\in B}M(y)P(y|x)
\end{equation}
holds. For any $y\in A_1$ let $y^p$ be the initial fragment of $y$ 
of maximal length such that $\frac{M(y^p)}{M(x)}>\frac{1}{(1-\mu)P(A|x)}$.
The set $\{y^p : y\in A_1\}$ is prefix free. Then by (\ref{super-1}) we have
\begin{eqnarray*}
1\ge\sum\limits_{y\in A_1}\frac{M(y^p)}{M(x)}P(y^p|x)>   \\
\frac{1}{(1-\mu)P(\tilde A|x)}\sum\limits_{y\in A_1}P(y^p|x)\ge
\frac{1}{(1-\mu)P(\tilde A|x)}P(\tilde A_1|x).
\end{eqnarray*}
From this we obtain $P(\tilde A_1|x)<(1-\mu)P(\tilde A|x)$. 
Define
$$
A'=A-\{y\in A : z\subseteq y\mbox { for some }\mbox z\in A_1\}.
$$
Then $P(\tilde A'|x)>\mu P(\tilde A|x)$. For any $y\in A'$ we have
$$
M(y^j)\le M(x)\frac{1}{(1-\mu)P(\tilde A|x)}
$$
for all $l(x)\le j\le(y)$. The result of the proposition follows
from inequality $d_P(x)\le\log M(x)$.
$\bigtriangleup$

\section{Appendix~2}

{\bf Method of cutting and stacking}.
An arbitrary measurable mapping of the a probability space 
into itself is called a transformation or a process.
A transformation $T$ preserves a measure $P$ if
$P(T^{-1}(A))=T(A)$ for all measurable subsets $A$ of the space.
A subset $A$ is called invariant with respect to $T$ if $T^{-1}A=A$.
A transformation $T$ is called ergodic if each invariant with respect
to $T$ subset $A$ has measure~0~or~1.

The simplest example of such transformation of the space 
$A^{\infty}$ of all infinite sequences, where $A=\{0,1,\dots, k-1\}$
is some finite alphabet, is the (left) shift $T$
defined by $(T\omega)_i=\omega_{i+1}$ for all $i=1,2,\dots$. If the shift $T$
preserves the measure $P$ then this measure is called stationary, i.e.
$$
P\{\omega : \omega_i=x_1,\dots,\omega_{i+k-1}=x_k\}=
P\{\omega : \omega_1=x_1,\dots,\omega_k=x_k\}
$$
for all positive integer numbers $i,k\ge 1$ and all $x_1,\dots, x_k$ equal
0~or~1. 

Recall some notions of symbolic dynamics.
We us consider the uniform measure $\lambda$ on the unit interval $[0,1)$
and a transformation $T$ of this interval. A partition is a sequence
pairwise disjoint subsets $\pi=(\pi_1,\dots,\pi_k)$ of the interval 
$[0,1)$ whose union is equal to this interval.
A transformation $T$ defines a measure on the set of all finite and
infinite words of the alphabet $A=\{0,1,\dots, k-1\}$ as follows
\begin{equation} \label{mes-1a}
P(a_1a_2\dots a_n)=\lambda\{\omega : 
\omega\in [0,1),\mbox T^i(\omega)\in\pi_{a_i},\mbox i=1,2,\dots,n\},
\end{equation}
where $a_1a_2\dots a_n$ is a sequence of letters from $A$.
The measure $P$ can be extended on all Borel subsets of
$A^{\infty}$ by a natural fashion \cite{Shi80}.
The measure $P$ defined by (\ref{mes-1a}) is stationary and ergodic
with respect to the left shift if and only if the transformation $T$ has
the same properties.

We use a cutting and stacking method of constructing of ergodic 
processes \cite{Shi91, Shi93}. 
Recall the main notions and properties of this method.
A column is a sequence $E=(L_1,\dots,L_h)$ of pairwise disjoint
subintervals of the unit interval of equal width;
$L_1$ is the base, $L_h$ is the top of the column,
${\hat E}=\cup_{i=1}^{h}L_i$ is the support of the
column, $w(E)=\lambda(L_1)$ is the width of the column, 
$h$ is the height of the column, 
$\lambda({\hat E})=\lambda(\cup_{i=1}^{h}L_i)$ is the measure of the column. 
Any column defines an algorithmically effective transformation $T$
which linearly transforms $L_j$ to $L_{j+1}$ for all $j=1,\dots, h-1$. 
This transformation $T$ is not defined outside all intervals of the column
and at all points of the top $L_h$ interval of this column.
Denote $T^0\omega=\omega$, $T^{i+1}\omega=T(T^i\omega)$.
For any $1\le j<h$ an arbitrary point $\omega\in L_j$ generates a finite
trajectory $\omega, T\omega,\dots, T^{h-j}\omega$.
A partition $\pi=(\pi_1,\dots,\pi_k)$ is compatible with a column $E$
if for each $j$ there exists an $i$ such that $L_j\subseteq\pi_i$.
This number $i$ is called the name of the interval $L_j$, and the
corresponding sequence of names of all intervals of the column is
called the name of the column $E$.
For any point $\omega\in L_j$, where $1\le j<h$, by $E$--name of
the trajectory $\omega, T\omega,\dots, T^{h-j}\omega$ we mean
a sequence of names of intervals $L_j,\dots, L_h$ from the column $E$. 
The length of this sequence is $h-j+1$.

A gadget is a finite collection of disjoint columns.
The width of the gadget $w(\Upsilon)$ is the sum of the widths of its
columns. A union of gadgets $\Upsilon_i$ with disjoint supports
is the gadget $\Upsilon=\cup\Upsilon_i$ whose columns are the columns 
of all the $\Upsilon_i$. The support of the gadget $\Upsilon$ is
the union $\hat\Upsilon$ of the supports of all its columns.
A transformation $T(\Upsilon)$ is associated with a gadget $\Upsilon$ if 
it is the union of transformations defined on all columns of $\Upsilon$.
With any gadget $\Upsilon$ the corresponding set of finite trajectories
generated by points of its columns is associated. By $\Upsilon$-name
of a trajectory we mean its $E$-name, where $E$ is that column 
of $\Upsilon$ to which this trajectory corresponds.
A gadget $\Upsilon$ extends a column $\Lambda$ if the support of
$\Upsilon$ extends the support of $\Lambda$, the transformation 
$T(\Upsilon)$ extends the transformation $T(\Lambda)$ and the partition
corresponding to $\Upsilon$ extends the partition corresponding to $\Lambda$.

The cutting and stacking operations that are common used will now be
defined.
The distribution of a gadget $\Upsilon$ with columns
$E_1,\dots,E_n$ is a vector of probabilities
$$
\left(\frac{w(E_1)}{w(\Upsilon)},\dots,\frac{w(E_n)}{w(\Upsilon)}\right).
$$
A gadget $\Upsilon$ is a copy of a gadget $\Lambda$ if they have the same
distribution and the corresponding columns have the same partition names.
A gadget $\Upsilon$ can be cut into $M$ copies of itself
$\Upsilon_i, i=1,\dots, M$, according to a given probability vector
$(\gamma_1,\dots,\gamma_n)$ by cutting each column
$E_i=(L_{i,j}: 1\le j\le h(E_i))$ (and its intervals) into disjoint
subcolumns $E_{i,m}=(L_{i,j,m}: 1\le j\le h(E_i))$ such that
$w(E_{i,m})=w(L_{i,j,m})=\gamma_m w(L_{i,j})$.
The gadget $\Upsilon_m=\{E_{i,m}:1\le i\le L\}$ is called the copy of
the gadget $\Upsilon$ of width $\gamma_m$. The action of the gadget
transformation $T$ is not affected by the copying operation.

Another operation is the stacking gadgets onto gadgets.
At first we consider the stacking of columns onto columns and
the stacking of gadgets onto columns.

Let $E_1=(L_{1,j}:1\le j\le h(E_1))$ and $E_2=(L_{2,j}:1\le j\le h(E_2))$
be two columns of equal width whose supports are disjoint.
The new column $E_1*E_2=(L_j:1\le j\le h(E_1)+h(E_2))$ is defined as
$L_j=L_{1,j}$ for all $1\le j\le h(E_1)$ and $L_j=L_{2,j-h(E_1)+1}$ for all
$h(E_1)\le j\le h(E_1)+h(E_2)$. 
Let a gadget $\Upsilon$ and a column $E$ have the same width, and their
supports are disjoint. A new gadget $E*\Upsilon$ is defined as follows.
Cut $E$ into subcolumns $E_i$ according to the distribution of the gadget
$\Upsilon$ such that $w(E_i)=w(U_i)$, where $U_i$ is the $i$-th column
of the gadget $\Upsilon$. Stack $U_i$ on the top of $E_i$ to get the new
column $E_i*U_i$. A new gadget consists of the columns $(E_i*U_i)$.

Let $\Upsilon$ and $\Lambda$ be two gadgets of the same width
and with disjoint supports. A gadget $\Upsilon*\Lambda$ is defined as
follows. Let the columns of $\Upsilon$ are $(E_i)$. Cut $\Lambda$ into
copies $\Lambda_i$ such that $w(\Lambda_i)=w(E_i)$ for all $i$.
After that, for each $i$ stack the gadget $\Lambda_i$ onto
column $E_i$, i.e. we consider a gadget $E_i*\Lambda_i$.
The new gadget is the union of gadgets $E_i*\Lambda_i$ for all $i$.
The number of columns of the gadget $\Upsilon*\Lambda$ is the product
of the number of columns of $\Upsilon$ on the number of columns of
$\Lambda$.

The $M$-fold independent cutting and stacking of a single gadget
$\Upsilon$ is defined by cutting $\Upsilon$ into $M$ copies $\Upsilon_i$,
$i=1,\dots,M$, of equal width and successively independently cutting
and stacking them to obtain $\Upsilon^{*(M)}=\Upsilon_1*\dots*\Upsilon_M$. 

{\bf Remark~1.}
Several examples of stationary measures constructed using cutting and stacking 
method are given in \cite{Shi91, Shi93}. We use in Section~\ref{non-robust-1}
a construction of a sequence of gadgets defining the uniform Bernoulli
distribution on trajectories generated by them. This sequence
is constructed using the following scheme. Let a partition
$\pi=(\pi_0,\pi_1)$ be given. Let also $\Delta$ be a gadget such that
its columns have the same width and are compatible with the partition
$\pi$. Let $\lambda(\hat\Delta\cap\pi_0)=\lambda(\hat\Delta\cap\pi_1)$.
Suppose that for some $M$ a gadget $\Delta'$ is constructed
from the gadget $\Delta$ by means of $M$--fold independent cutting
and stacking and $P$ be a measure on trajectories of the gadget $\Delta'$
defined by (\ref{mes-1a}). Then by the method of cutting and stacking
$P(x)=2^{-l(x)}\lambda(\hat\Delta)$ for the trajectory $x$ of any
point from the support of $\hat\Delta'$.

A sequence of gadgets $\{\Upsilon_m\}$ is complete if
\begin{itemize}
\item{}
$\lim\limits_{m\to\infty} w(\Upsilon_m)=0$;
\item{}
$\lim\limits_{m\to\infty} \lambda({\hat\Upsilon}_m)=1$;
\item{}
$\Upsilon_{m+1}$ extends $\Upsilon_m$ for all $m$.
\end{itemize}
Any complete sequence of gadgets $\{\Upsilon_s\}$ determines a transformation
$T=T\{\Upsilon_s\}$ which is defined on interval $[0,1)$ almost surely.

By definition $T$ preserves the measure $\lambda$. In \cite{Shi91} 
and \cite{Shi93} the conditions sufficient a process $T$ to be ergodic 
were suggested. Let a gadget $\Upsilon$ is constructed by cutting and 
stacking from a gadget $\Lambda$.
Let $E$ be a column from $\Upsilon$ and $D$ be a column from $\Lambda$. 
Then ${\hat E}\cap{\hat D}$ is defined as the union of subcolumns from
$D$ of width $w(E)$ which were used for construction of $E$. 

Let $0<\epsilon<1$.
A gadget $\Lambda$ is $(1-\epsilon)$-well-distributed in $\Upsilon$ if
\begin{equation} 
\sum_{D\in\Lambda}\sum_{E\in\Upsilon}|\lambda({\hat E}\cap{\hat D})-
\lambda({\hat E})\lambda({\hat D})|<\epsilon.
\end{equation}
We will use the following two lemmas.
\begin{lemma} \label{well-def}
(\cite{Shi91}, Corollary 1), (\cite{Shi93}, Theorem A.1).
Let $\{\Upsilon_n\}$ be a complete sequence of gadgets and for each $n$
the gadget $\{\Upsilon_n\}$ is $(1-\epsilon_n)$-well-distributed in 
$\{\Upsilon_{n+1}\}$, where $\epsilon_n\to 0$. Then $\{\Upsilon_n\}$
defines the ergodic process.
\end{lemma}
\begin{lemma} \label{M-fold}
(\cite{Shi93}, Lemma 2.2).
For any $\epsilon>0$ and any gadget $\Upsilon$ there is an $M$ such that
for each $m\ge M$ the gadget $\Upsilon$ is $(1-\epsilon)$-well-distributed
in the gadget $\Upsilon^{*(m)}$ constructed from $\Upsilon$ by
$\mbox m$-fold independent cutting and stacking.
\end{lemma}

}

\end{document}